\documentclass[10pt,A4paper,twocolumn,aps,pra, superscriptaddress,longbibliography, nofootinbib]{revtex4-2}

\usepackage{amsmath}
\usepackage{graphicx}
\usepackage{color}
\usepackage[usenames,dvipsnames,svgnames,table]{xcolor}
\usepackage[hidelinks]{hyperref}
\usepackage{float}
\usepackage{braket}
\usepackage{bbm}
\usepackage{multibib}
\newcites{meth}{Methods References}

\usepackage[capitalise]{cleveref}
\Crefname{figure}{Fig.}{Figs.}
\Crefname{table}{Tab.}{Tabs.}
\usepackage{siunitx}
\usepackage{amssymb}

\usepackage{rotating}
\usepackage{tabularx}
\usepackage{cellspace}
\usepackage{booktabs}
\usepackage{graphics}
\usepackage{newfloat}
\usepackage{soul}

\usepackage{xr}
\usepackage[T1]{fontenc}

\makeatother

\usepackage{chemformula}

\begin{document}

\begin{abstract}

The dominant contribution to the energy relaxation of state-of-the-art superconducting qubits is often attributed to their coupling to an ensemble of material defects which behave as two-level systems. These defects have varying microscopic characteristics which result in a large range of observable defect properties such as resonant frequencies, coherence times and coupling rates to qubits $g$. 
Here, we investigate strategies to mitigate losses to the family of defects that strongly couple to qubits ($g/2\pi\ge\SI{0.5}{MHz}$). Such strongly coupled defects are particularly detrimental to the coherence of qubits and to the fidelities of operations relying on frequency excursions, such as flux-activated two-qubit gates.
To assess their impact, we perform swap spectroscopy on 92 frequency-tunable qubits and quantify the spectral density of these strongly coupled modes. We show that the frequency configuration of the defects is rearranged by warming up the sample to room temperature, whereas the total number of defects on a processor tends to remain constant. We then explore methods for fabricating qubits with a reduced number of strongly coupled defect modes by systematically measuring their spectral density for decreasing Josephson junction dimensions and for various surface cleaning methods. Our results provide insights into the properties of strongly coupled defect modes and show the benefits of minimizing Josephson junction dimensions to improve qubit properties.

\end{abstract}

\date{July 26, 2024}

\author{Dante~Colao~Zanuz}
\affiliation{Department of Physics, ETH Zurich, CH-8093 Zurich, Switzerland}
\affiliation{Quantum Center, ETH Zurich, CH-8093 Zurich, Switzerland}

\author{Quentin~Ficheux}
\affiliation{Department of Physics, ETH Zurich, CH-8093 Zurich, Switzerland}

\author{Laurent~Michaud}
\affiliation{Department of Physics, ETH Zurich, CH-8093 Zurich, Switzerland}

\author{Alexei~Orekhov}
\affiliation{Department of Physics, ETH Zurich, CH-8093 Zurich, Switzerland}

\author{Kilian~Hanke}
\affiliation{Department of Physics, ETH Zurich, CH-8093 Zurich, Switzerland}

\author{Alexander~Flasby}
\affiliation{Department of Physics, ETH Zurich, CH-8093 Zurich, Switzerland}
\affiliation{Quantum Center, ETH Zurich, CH-8093 Zurich, Switzerland}
\affiliation{ETH Zurich - PSI Quantum Computing Hub, Paul Scherrer Institute, CH-5232 Villigen, Switzerland}

\author{Mohsen~Bahrami~Panah}
\affiliation{Department of Physics, ETH Zurich, CH-8093 Zurich, Switzerland}
\affiliation{Quantum Center, ETH Zurich, CH-8093 Zurich, Switzerland}
\affiliation{ETH Zurich - PSI Quantum Computing Hub, Paul Scherrer Institute, CH-5232 Villigen, Switzerland}

\author{Graham~J.~Norris}
\affiliation{Department of Physics, ETH Zurich, CH-8093 Zurich, Switzerland}
\affiliation{Quantum Center, ETH Zurich, CH-8093 Zurich, Switzerland}

\author{Michael~Kerschbaum}
\affiliation{Department of Physics, ETH Zurich, CH-8093 Zurich, Switzerland}
\affiliation{Quantum Center, ETH Zurich, CH-8093 Zurich, Switzerland}
\affiliation{ETH Zurich - PSI Quantum Computing Hub, Paul Scherrer Institute, CH-5232 Villigen, Switzerland}

\author{Ants~Remm}
\affiliation{Department of Physics, ETH Zurich, CH-8093 Zurich, Switzerland}

\author{François~Swiadek}
\affiliation{Department of Physics, ETH Zurich, CH-8093 Zurich, Switzerland}
\affiliation{Quantum Center, ETH Zurich, CH-8093 Zurich, Switzerland}

\author{Christoph~Hellings}
\affiliation{Department of Physics, ETH Zurich, CH-8093 Zurich, Switzerland}
\affiliation{Quantum Center, ETH Zurich, CH-8093 Zurich, Switzerland}

\author{Stefania~Lazăr}
\affiliation{Department of Physics, ETH Zurich, CH-8093 Zurich, Switzerland}

\author{Colin~Scarato}
\affiliation{Department of Physics, ETH Zurich, CH-8093 Zurich, Switzerland}
\affiliation{Quantum Center, ETH Zurich, CH-8093 Zurich, Switzerland}

\author{Nathan~Lacroix}
\affiliation{Department of Physics, ETH Zurich, CH-8093 Zurich, Switzerland}
\affiliation{Quantum Center, ETH Zurich, CH-8093 Zurich, Switzerland}

\author{Sebastian~Krinner}
\affiliation{Department of Physics, ETH Zurich, CH-8093 Zurich, Switzerland}

\author{Christopher~Eichler}
\affiliation{Department of Physics, ETH Zurich, CH-8093 Zurich, Switzerland}

\author{Andreas~Wallraff}
\affiliation{Department of Physics, ETH Zurich, CH-8093 Zurich, Switzerland}
\affiliation{Quantum Center, ETH Zurich, CH-8093 Zurich, Switzerland}
\affiliation{ETH Zurich - PSI Quantum Computing Hub, Paul Scherrer Institute, CH-5232 Villigen, Switzerland}

\author{Jean-Claude~Besse}
\affiliation{Department of Physics, ETH Zurich, CH-8093 Zurich, Switzerland}
\affiliation{Quantum Center, ETH Zurich, CH-8093 Zurich, Switzerland}
\affiliation{ETH Zurich - PSI Quantum Computing Hub, Paul Scherrer Institute, CH-5232 Villigen, Switzerland}

\title{Mitigating Losses of Superconducting Qubits Strongly Coupled to Defect Modes}

\maketitle

\section*{Introduction} \label{sec:introduction}

\begin{figure*}[]
\centering
	\includegraphics[width=1\linewidth]{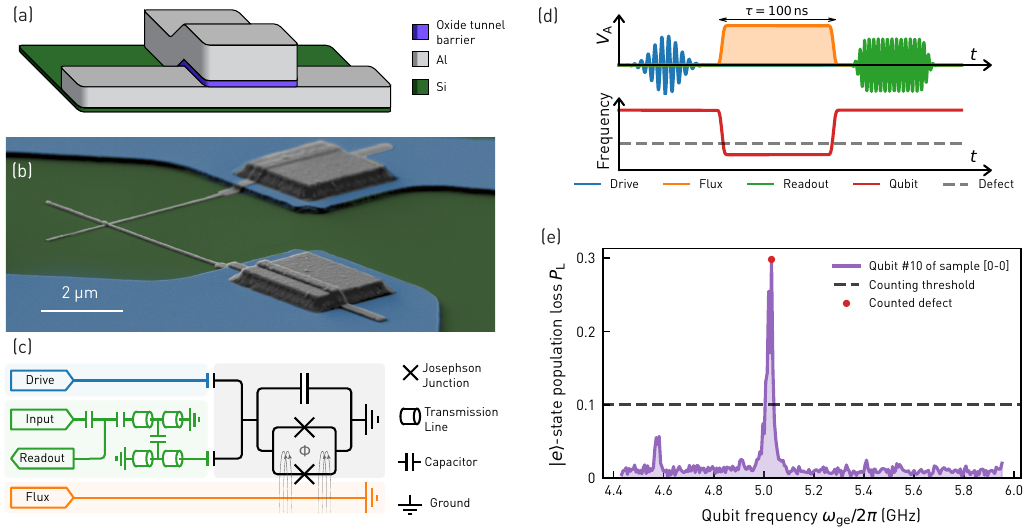}

	\caption{(a) 3D schematic of a Josephson junction. (b) False-colored scanning electron micrograph of a Josephson junction. Green indicates silicon, blue niobium and gray  aluminum.
	(c) Equivalent circuit diagram of the on-chip connectivity of the samples. (d) Pulse sequence of swap-spectroscopy with defect modes and the qubit transition frequency resulting from the shown flux-control pulse. (e) Swap-spectrum of one representative qubit depicting the population loss $P_\mathrm{L}$ of the excited state after \SI{100}{\ns} of interaction time versus the qubit interaction frequency $\omega_{ge}/2\pi$ (purple curve). The counting threshold is represented by the black dashed line and the red dot corresponds to the only counted defect for the displayed data.}
	\label{fig:fig1} 
		
\end{figure*}

The performance of current superconducting quantum processors suffers greatly from decoherence of qubits that store and process quantum information~\cite{Siddiqi2021,Spiecker2023,Kjaergaard2020a}.  
In state-of-the-art transmon qubits, the dominant contribution to energy relaxation is often attributed to dielectric losses stemming from material defects present in the sample, which behave as two-level systems (TLSs)~\cite{Wang2015n,Mueller2019}. 
Qubits interact with such defects via the coupling between the electrical field of the qubit, $\mathbf{E}$, and the defect electric dipole moment, $\mathbf{p}$, with an overall coupling rate $g=\mathbf{p}\cdot \mathbf{E} / \hbar$. 
Energy, once transferred to the TLS, can be dissipated as heat via their coupling to phonons~\cite{Muller2009}. 
Defects are particularly dense in amorphous oxides~\cite{Woods2019, Altoe2022}, which are abundant in superconducting circuits as they form at the sample's interfaces to air. 
These regions tend to contribute significantly to the losses of superconducting qubits due to their large areas and high loss tangents~\cite{Woods2019}.
However, the electric field amplitudes at these interfaces are typically as weak as a few \SI{}{V/m}, which leads to small coupling rates $g/2\pi$ of hundreds of \SI{}{kHz} for most types of material defects~\cite{Lisenfeld2019,Bilmes2020}.
This underpins the common understanding that the coherence of qubits is ultimately limited by their interaction with an ensemble of weakly coupled defect modes~\cite{Mueller2015,Neill2013,Burnett2014,Bilmes2020}.

In addition to weakly coupled defects, coherent defects with coupling rates that can be hundreds of times larger are observed~\cite{Simmonds2004,Simmonds2009,Cole2010,Lisenfeld2019,Mamin2021,Klimov2018}. 
When resonant with a fixed-frequency qubit, strongly coupled defect modes can render a qubit practically unusable. 
Defects are also harmful in frequency-tunable qubit architectures, where they restrict the choice of idling frequencies~\cite{Sung2021a} and impair the fidelities of flux-enabled operations, such as two-qubit gates~\cite{DiCarlo2009, Krinner2022}, flux-pulse-assisted readout~\cite{Swiadek2023} and reset~\cite{Lacroix2023}.
Strongly coupled modes are suspected to arise from defects located in regions with particularly intense electric fields, such as at the Josephson junctions~\cite{Simmonds2009,Lisenfeld2019,Bilmes2022}.
There, electric fields of several \SI{}{kV/m} are expected from the zero-point fluctuations of charge carriers $n_{\mathrm{zpf}}$ for typical transmon parameters:
\begin{equation}
    |\mathbf{E_{\mathrm{zpf}}}|\approx\frac{2\,e\,n_{\mathrm{zpf}}}{C\,d},
\end{equation}
where $e$ is the electron charge, $C$ is the qubit's total capacitance and $d$ is the tunnel barrier thickness. The thickness of a Josephson junction barrier is often inhomogeneous~\cite{Zeng2015}, which broadens the range of coupling rates between qubits and defects. 

In a conventional \ch{Al}/\ch{AlO_x}/\ch{Al} Josephson junction for superconducting circuits, as shown in Fig.~\ref{fig:fig1}(a,b), two particular regions are believed to host the highest concentration of defects:
(1) the \textit{aluminum oxide tunnel barrier}, where not only the electric field is the strongest, but also the amorphous structure of the oxide is expected to host multiple material defects~\cite{DuBois2013,Zeng2016,Holder2013,Gordon2014}; 
and (2) the \textit{junction-substrate interface}, 
where the junction fabrication based on a liftoff process may leave organic residues and contaminants~\cite{Pop2012a} on a roughened silicon substrate~\cite{Quintana2014}. 
This can ultimately influence the morphology of aluminum grains grown on top of the substrate and introduce additional TLSs~\cite{Zeng2015a, Fritz2018}.
 
Since large-scale quantum computations require improvements in gate fidelities, coherence times, and fabrication yield, there is a pressing need to reduce the number of strongly coupled defect modes. 
In this manuscript, we introduce metrics to quantify strongly coupled defect modes in our samples and evaluate them for 92 frequency-tunable qubits. We first investigate the dynamics of the defect modes and analyze their properties when the sample is warmed up to room temperature and cooled back down to approximately \SI{10}{mK}. We also study fabrication and design approaches to mitigate the formation of defects by studying the two previously defined critical regions of the Josephson junctions: we explore the mitigation of losses at the junction oxide barrier by reducing the sizes of the junctions and we 
address the cleanliness of the vicinity of the junctions by performing different cleaning treatments before the junction deposition.

\section{Characterizing defect modes}
\label{sec:characterization}

\begin{figure*}[]
\centering
	\includegraphics[width=1\linewidth]{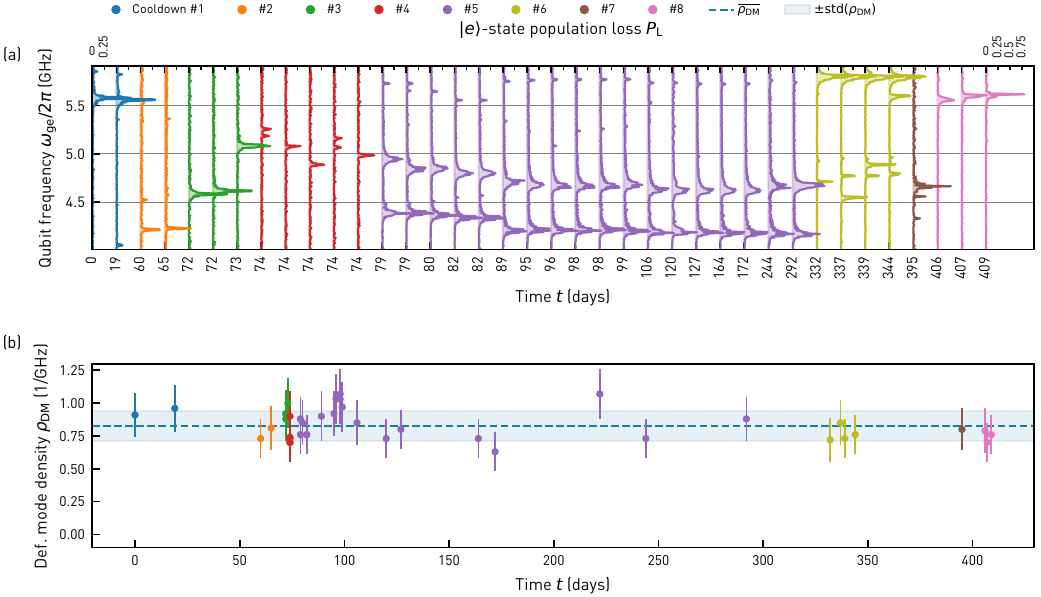}
	\caption{(a) Changes in the swap spectrum of a single exemplary qubit from sample [0-0] over time and thermal cycles.
	The colors identify the periods during which the sample was kept at \SI{10}{mK}. (b) Spectral defect mode density extracted for all 17 qubits in the same device.
 The dashed line $\overline{\rho_{\mathrm{DM}}}$ and filled region correspond respectively to the averaged $\rho_{\mathrm{DM}}$ from all data and their standard deviation.
    }
	\label{fig:fig2}
\end{figure*}

The coupling between a TLS and a qubit induces a quantifiable energy exchange~\cite{Martinis2005,Ashhab2006} which we leverage to characterize defects. 
Here, we investigate defect signals in transmons designed with parameters relevant for quantum information processing applications, including quantum error correction~\cite{Krinner2022}.
Each transmon is connected to a Purcell-filtered readout resonator, as well as individual flux and drive lines, as depicted in the circuit diagram in Fig.~\ref{fig:fig1}(c). 
We fabricate the samples on a silicon substrate. The large micrometer-scale features are made of niobium while the Josephson junctions are defined in an \ch{Al}/\ch{AlO_x}/\ch{Al} structure (see details of the fabrication methods in Appendix~\ref{sec:supplemental_fab}).
We label chip $c$ from wafer $w$ as [$w$-$c$] and tabulate individual chip details in Appendix~\ref{sec:supplemental-table}.

We initiate the measurement protocol by applying a constant current bias to the flux line such that the magnetic flux $\Phi$ threading the qubit SQUID loop tunes the transition frequency of the qubit to a \textit{sweet spot}~\cite{Koch2007}, where it is first-order insensitive to flux noise.
The defect modes are then probed via \emph{swap-spectroscopy}~\cite{Cooper2004, Shalibo2010, Lisenfeld2016, Lisenfeld2019} as depicted schematically in Fig.~\ref{fig:fig1}(d). A microwave pulse injected into the drive line (blue) excites the qubit from its ground state $\ket{g}$ to its first excited state $\ket{e}$. A flux pulse (orange) of amplitude $V_A$ then induces the frequency excursion of the qubit to a target frequency, at which the qubit remains for an interaction time of $\tau = \SI{100}{\ns}$, a time scale similar to
the typical duration of two-qubit controlled-phase gates~\cite{Strauch2003,Negirneac2021}.  
Finally, the flux pulse ends and the qubit returns to its idling frequency, at which standard dispersive readout (green) is performed.
We define the $\ket{e}$-state population loss as $P_\mathrm{L} = 1 - P_{\ket{e}}$, where $P_{\ket{e}}$ is the qubit population measured in the $\ket{e}$ state at the end of the experiment. We record $P_\mathrm{L}$ for varying flux-pulse amplitudes to probe the entire frequency range accessible by tuning the qubit.

The frequency dependence of the population loss, which we refer to as the \emph{swap spectrum}, generally features two distinct regimes, as exemplified in Fig.~\ref{fig:fig1}(e): (1) frequencies at which the population loss is dictated by the qubit's loss to a background bath of weakly-coupled defects;  and (2) frequencies where the population loss peaks which we attribute to the interaction with individual strongly coupled defect modes.
To quantify the number of these modes per frequency bandwidth, we define the \emph{spectral density of strongly coupled defect modes} $\rho_{\mathrm{DM}}$.
This quantity corresponds to the number of loss peaks above a set threshold per unit of analyzed bandwidth, as depicted in Fig.~\ref{fig:fig1}(e). 
We choose a threshold of 10\% of population loss $P_\mathrm{L}$ in an interaction time of $\tau =\SI{100}{ns}$ so that the threshold is sufficiently above the background loss level. 
This threshold translates into a condition of identifying only defect modes with 
$g/2\pi \ge \mathrm{arccos}(1-2\,P_\mathrm{L})/(4\pi\,\tau) = \SI{0.5}{MHz}$, where the bound corresponds to a purely coherent exchange of energy.
We analyze the robustness of our results to alternative threshold values in Appendix~\ref{sec:supplemental-threshold}, finding no impact on our conclusions.
The data analysis is described in Appendix~\ref{sec:supplemental-algorithm}.

\section{Drift of defect mode frequencies and the effect of thermal cycling}

To investigate the change in the defect mode configurations over time, we tracked the swap spectra of 17 qubits measured throughout a period of 409 days on a distance-three surface code device~\cite{Krinner2022}, see Fig.~\ref{fig:fig2}(a) for a typical qubit. 
At \SI{10}{mK}, we repeatedly observe defect mode frequencies drifting a few \SI{}{MHz/day}, consistent with spectral diffusion arising from the interaction of the defect modes with multiple fluctuators~\cite{Black1977,Mueller2015,Schloer2019}. 
In some cases, we observe substantial reconfiguration of the defect mode frequencies (\textit{e.g.} between days 72 and 73), which we attribute to the mutual interaction between high-energy coherent TLSs~\cite{Lisenfeld2015} or to background ionizing radiation~\cite{Thorbeck2023a}.

The sample has been thermally cycled to room-temperature five times and partially to \SI{10}{K} and \SI{60}{\K} before cooldowns 4 and 6, respectively. Each cooldown is marked by a color in Fig.~\ref{fig:fig2}(a). We observed that all thermal cycles reorganized the defect mode configuration, often increasing or decreasing the number of peaks observed within the measured range. 
This observation agrees with the energetic cost of defect reorganization lying well below the thermal energy at room temperature~\cite{Shalibo2010,Burnett2019}.

Despite changes in the swap spectra, the spectral defect mode density $\rho_{\mathrm{DM}}(\mathrm{qb}_1,\cdots ,\mathrm{qb}_{17})$ of all 17 qubits in the chip remained approximately unchanged across all the thermal cycles, as shown in Fig.~\ref{fig:fig2}(b). As a consequence, while thermal cycles can be exploited to rearrange the defect mode configuration of a sample, it becomes exponentially less likely to reach adequate defect mode configurations for all qubits as the number of qubits in the sample grows larger. 
It is therefore essential to minimize the formation of strongly-coupled defect modes in order to scale up superconducting quantum processors to larger qubit numbers.

\section{Losses within the barrier oxide of Josephson junctions}

\begin{figure}[]
\centering
\includegraphics{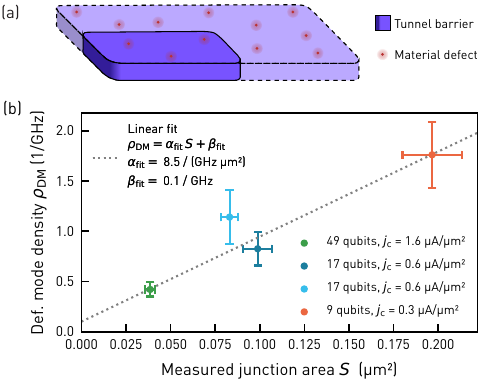}
\caption{
(a) Schematic representing a uniform distribution of material defects in a Josephson junction. 
At constant spatial density of defects, junctions with smaller overlap areas (dark purple) display a lower number of defects as compared to junctions with a larger overlap area (light purple).
(b) Cumulative spectral density of strongly coupled defect modes $\rho_{\mathrm{DM}}$ for qubits with the indicated total junction area $S$. The areas have been measured using SEM. The dashed line is a linear fit. Samples originate from wafers 0 (dark blue), 1 to 3 (green), 4 (cyan), and 5 (orange).} 
\label{fig:fig3}
\end{figure}

To avoid the formation of strongly coupled defect modes, we explore improvements in the fabrication of the Josephson junction, where most of the strongly coupled defects are expected to reside. 
We specifically investigate the reduction of the number of defect modes within the tunnel barrier by reducing the Josephson junction area, as shown in Fig.~\ref{fig:fig3}(a). This strategy has previously been shown to be effective in reducing the number of defect modes in phase qubits~\cite{Simmonds2009} and in parasitic junctions~\cite{Bilmes2022}. In this work, we focus on defects in the Josephson junctions that form the qubit. We avoid the formation of parasitic junctions by using a Manhattan style architecture~\cite{Potts2001, Costache2012, Kreikebaum2019} for all samples except [0-0]. Sample [0-0] was fabricated with Dolan-style junctions~\cite{Dolan1977} for the work in Ref.~\cite{Krinner2022}.

Reducing the overlap area of the main Josephson junctions reduces the Josephson energy ($E_\mathrm{J}$), which we compensate by fabricating junctions with larger critical current densities $j_\mathrm{c}$ such that the qubit transition frequency remains constant. This is achieved by adjusting the barrier oxidation times and pressures to three configurations for a mixture of 15\% \ch{O2} in \ch{Ar}: 
[\SI{3}{min},~\SI{3}{Torr}] for $j_\mathrm{c} = $~\SI{1.6}{\micro \A \per \micro m^2},
[\SI{12}{min},~\SI{10}{Torr}] for $j_\mathrm{c} = $~\SI{0.6}{\micro \A \per \micro m^2} and 
[\SI{60}{min},~\SI{50}{Torr}] for $j_\mathrm{c} = $~\SI{0.3}{\micro \A \per \micro m^2}. 
Lower times and pressures for the oxidation process lead to thinner tunnel barriers. 
However, due to the exponential nature of the tunneling process, the critical current of the Josephson junctions can be changed by orders of magnitudes with small changes in the barrier thickness~\cite{Zeng2015}.

We analyzed a total of 92 qubits with junctions in three area ranges, with total junction areas $S=S_{\mathrm{j1}}+S_{\mathrm{j2}}$ ranging from \SI{0.034}{\micro \m^2} to \SI{0.22}{\micro \m^2}, where $S_{\mathrm{j}}$ are the areas of the individual junctions of the SQUID loop as determined by scanning electron microscopy (SEM). We required large qubit statistics to compensate for the relatively rare resonant defect modes and to attribute statistically significant expectation values of defect mode density $\rho_{\mathrm{DM}}$ to each junction size.
Each qubit provided, on average, an explorable frequency bandwidth of around \SI{1.6}{GHz}.

We observed that the spectral density of strongly coupled defect modes $\rho_{\mathrm{DM}}$ decreased as we made the Josephson junction areas smaller, as shown in Fig.~\ref{fig:fig3}(b). The trend supports linear scaling, in agreement with observations reported in the literature~\cite{Bilmes2022, Mamin2021}. 
These results suggest that Josephson junctions with small areas and high critical current densities can be beneficial for superconducting qubits as they minimize the incidence of resonant defect modes. 
This scaling also introduces evidence that the observed strongly coupled defect modes arise from TLSs inside the tunnel barrier, such as oxygen vacancies~\cite{Zeng2016} and hydrogen contaminants~\cite{Holder2013, Gordon2014}, which are expected to increase in number as the area of tunnel barriers grow larger. 
We note however, that our results cannot distinguish whether the defect mode densities are affected primarily by the different junction sizes or by the different critical current densities as we prioritized analyzing qubits with similar properties ($E_\mathrm{J}$ and charging energy $E_\mathrm{C}$) and scan for defects within a similar frequency bandwidth.   

While the reduction of the junction area lowers the spectral defect mode density, this strategy alone cannot make the defect mode density arbitrarily small, as it is constrained by lithographic precision and by the frequency targeting at small junction areas~\cite{Osman2023}. 
These limitations highlight the importance of future work on reducing the incidence of material defects in the tunnel barrier with optimized fabrication procedures~\cite{Fritz2018,Fritz2019} or with alternative barrier materials such as crystalline insulators~\cite{Lee2019g}.

\section{Losses in the vicinity of Josephson junctions}

\begin{figure}[]
\centering
\includegraphics[width=1\linewidth]{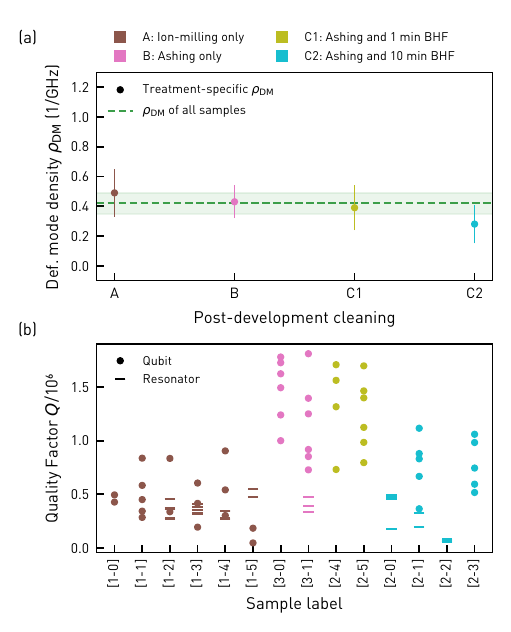}
\caption{(a) Spectral defect mode density $\rho_{\mathrm{DM}}$ for devices cleaned with the indicated post-development treatments.
The green dashed line and the green shaded area correspond respectively to the defect mode density and its uncertainty when the defect mode density is computed for all samples cleaned with any of the four treatments.
(b) Quality factors $Q$ of the measured individual qubits (dots) for the different analyzed chips and treatments: 
ion-milling (A, brown), ashing (B, pink), ashing and \SI{1}{min} of BHF (olive) and ashing and \SI{10}{min} of BHF (cyan).
Internal quality factor of test resonators included in the samples are indicated by dashes. Details for the individual samples are described in Appendix \ref{sec:supplemental-table}.
}
\label{fig:fig4}
\end{figure}

Finally, we investigate defects associated with the interfaces of the junction to the substrate and the base layer metallization. In typical fabrication processes of superconducting qubits, the junctions are patterned and deposited in a liftoff process known as shadow evaporation~\cite{Dolan1977}. Liftoff processes, however, are known to leave a thin layer of resist residues below the metalization, which can induce loss~\cite{Quintana2014} and render the junction properties unstable over time~\cite{Pop2012a, Koppinen2007}. 
Additionally, any processing in the vicinity of the junction can introduce oxides, charges, hydrates, impurities and damage to the structure~\cite{Dunsworth2017}, which can modify the morphology of the grown aluminum oxide and exacerbate defects both in and outside of the tunnel barrier~\cite{Fritz2018,Fritz2019}.

To investigate the impact of this interface on the spectral defect-mode density, we analyzed different sets of cleaning treatments applied after the electron-beam lithography (EBL) resist has been developed and prior to the aluminum deposition. At this step of the process,
the cleaning treatments predominantly affect the area underneath the to-be-fabricated Josephson junction. 

The four treatments tested aim first at removing the resist residue in the intended junction area, for which we investigated (A) \textit{in-situ argon ion milling},  
which removes organics and oxides, but is known to damage the substrate structure~\cite{VanDamme2023} and (B) \textit{ex-situ oxygen plasma ashing} to descum the substrate~\cite{Dunsworth2017}. Oxygen plasmas, however, also oxidize the substrate and base metallization, which can in turn increase losses. 
To eliminate this oxide layer, we followed the oxygen plasma ashing by a \textit{buffered hydrofluoric acid} (BHF) treatment~\cite{Chu2016b, Blok2021, Muthusubramanian2024, Moskalev2023} for the durations of (C1) \SI{1}{min} and (C2) \SI{10}{min}. The BHF treatments are expected to remove the oxides exclusively from silicon in the short treatment C1 and to remove the oxides both from silicon and from the niobium base metallization in the longer treatment C2~\cite{Altoe2022}.

To benchmark the four post-development processes, we have characterized a total of 49 qubits. 
We observe that the spectral density of the strongly coupled defect modes is not significantly affected by the different post-development treatments, see Fig.~\ref{fig:fig4}(a). A detailed analysis of a potential weak correlation warrants further study.

In contrast to the effects on $\rho_\mathrm{DM}$, the cleaning treatments had a significant influence on the qubit quality factors $Q=\omega_{\mathrm{qb}} \, T_1$, measured when their transition frequency was tuned to about $\omega_{\mathrm{qb}}/2\pi=\SI{6.0}{} \, \pm\, \SI{0.5}{GHz}$. We observe [see Fig.~\ref{fig:fig4}(b)] that the devices which have been treated with ion-milling presented lower mean quality factors, of 
$Q_{\mathrm{A}}= (0.46 \pm 0.23) \times 10^6$, when compared to the qubits treated with only oxygen plasma ashing, for which 
$Q_{\mathrm{B}}=(1.32 \pm 0.36) \times10^6$. This trend is consistent with results from Dunsworth \textit{et al.}~\cite{Dunsworth2017}. 
When the oxygen plasma ashing is followed by an additional BHF treatment, however, we find that the mean quality factors decrease with increasing duration of the BHF treatment, as 
$Q_{\mathrm{C1}}= (1.28 \pm 0.34) \times 10^6$ and $Q_{\mathrm{C2}}= (0.78 \pm 0.23) \times 10^6$.
We hypothesize this is due to the diffusion of the BHF through the resist stack leading to both damage in the aluminum structures and a layer of fluoride residues which were increasingly more apparent in SEM images taken on samples exposed to longer treatments.
Consequently, a more appropriate resist stack may need to be explored in the future for such treatments.

To rule out correlations between the quality of the base layer metallization and the observed trend of qubit quality factors from junction cleaning treatments, we also measured the internal quality factors of test resonators included on the qubit samples, which were consistently 
$Q_{\mathrm{int}}=(0.33 \pm 0.14) \times 10^6$ for the different chips and cleaning treatments.

As the overall quality factor of qubits is understood to be limited by the weakly coupled bath of defects~\cite{Mueller2015, Neill2013, Burnett2014},  the observations indicate that the weakly coupled and strongly coupled modes are physically distinguishable either in nature or in their location on the device. This is consistent with the strongly coupled defects being located at the junction barrier whereas the weakly coupled defects are located at the remaining surfaces of the samples where the electric field is weaker in comparison to that in the junction barrier.

\section{Conclusion}
Superconducting qubits can couple strongly to defects in the sample. 
When interacting with a qubit, such defects impair the fidelities of qubit operations,  induce readout errors and lower qubit coherence times.  
As experiments with superconducting quantum processors work with an increasing number of qubits, encountering qubits with detrimental strongly coupled defect mode configurations becomes more likely. 

To address this scaling challenge, we quantified the number of strongly coupled defect modes per unit of analyzed bandwidth over time and for Josephson junctions with different sizes and fabrication processes. 
We observed that while the defect mode configuration of a qubit changes substantially when the sample is thermally cycled, the overall number of defect modes remained approximately constant in 17 qubits analyzed throughout eight cooldowns. 
We additionally observed in a study with 92 qubits that the measured spectral defect mode density scales with the area of the Josephson junctions when $E_\mathrm{J}$ is kept constant. 
This provides a simple strategy to reduce the number of strongly coupled defect modes by reducing the junction area.
Lastly, we studied how procedures to clean the interface below the Josephson junctions affect the number of observed defect modes and the qubit quality factors. 
We observed that the spectral defect mode density was not significantly changed for the treatments we explored, while the quality factors of the samples were improved by about a factor three. 
The lack of correlation between strongly coupled modes and qubit quality factors suggests that the weakly and strongly coupled defects are physically distinguishable either in nature or in location on the device. 

Our observations are consistent with the strongly coupled defect modes originating from material defects within the tunnel barrier, where the electric field is most intense, while the weakly coupled defect modes originate from regions with relatively weaker electric field, such as at the vicinity of the Josephson junction and metal-air interface of the device.

\section*{Acknowledgments}
The authors acknowledge financial support by 
the Office of the Director of National Intelligence (ODNI), Intelligence Advanced Research Projects Activity (IARPA), via the U.S. Army Research Office grants W911NF-16-1-0071 and W911NF-23-2-0212, 
by the National Centre of Competence in Research Quantum Science and Technology (NCCR QSIT), 
a research instrument of the Swiss National Science Foundation (SNSF), 
by the SNSF R'equip grant 206021-170731, 
by the Baugarten Foundation and the ETH Zurich Foundation, 
and by ETH Zurich. 
The views and conclusions contained herein are those of the authors and should not be interpreted as necessarily representing the official policies or endorsements, either expressed or implied, of the ODNI, IARPA, or the U.S. Government.

\section*{Author contribution}
DCZ, JCB, QF and AW conceived the project. QF, AO, DCZ, KH, CH, CS and NL acquired the data. DCZ, QF, LM and AO contributed to the analysis. CH and SL contributed to the measurement setup. DCZ, JCB and SK developed the Josephson junction fabrication process. AR and FS designed the samples. DCZ, JCB, AR, AF, MK, GJN, MBP and SK fabricated the samples. JCB, QF, CH, CE and AW supervised the project. DCZ, QF and JCB wrote the manuscript with inputs from all authors.


\begin{appendix}
\section{Device fabrication}
\label{sec:supplemental_fab}

\begin{figure}[]
\centering
\includegraphics[width=1\linewidth]{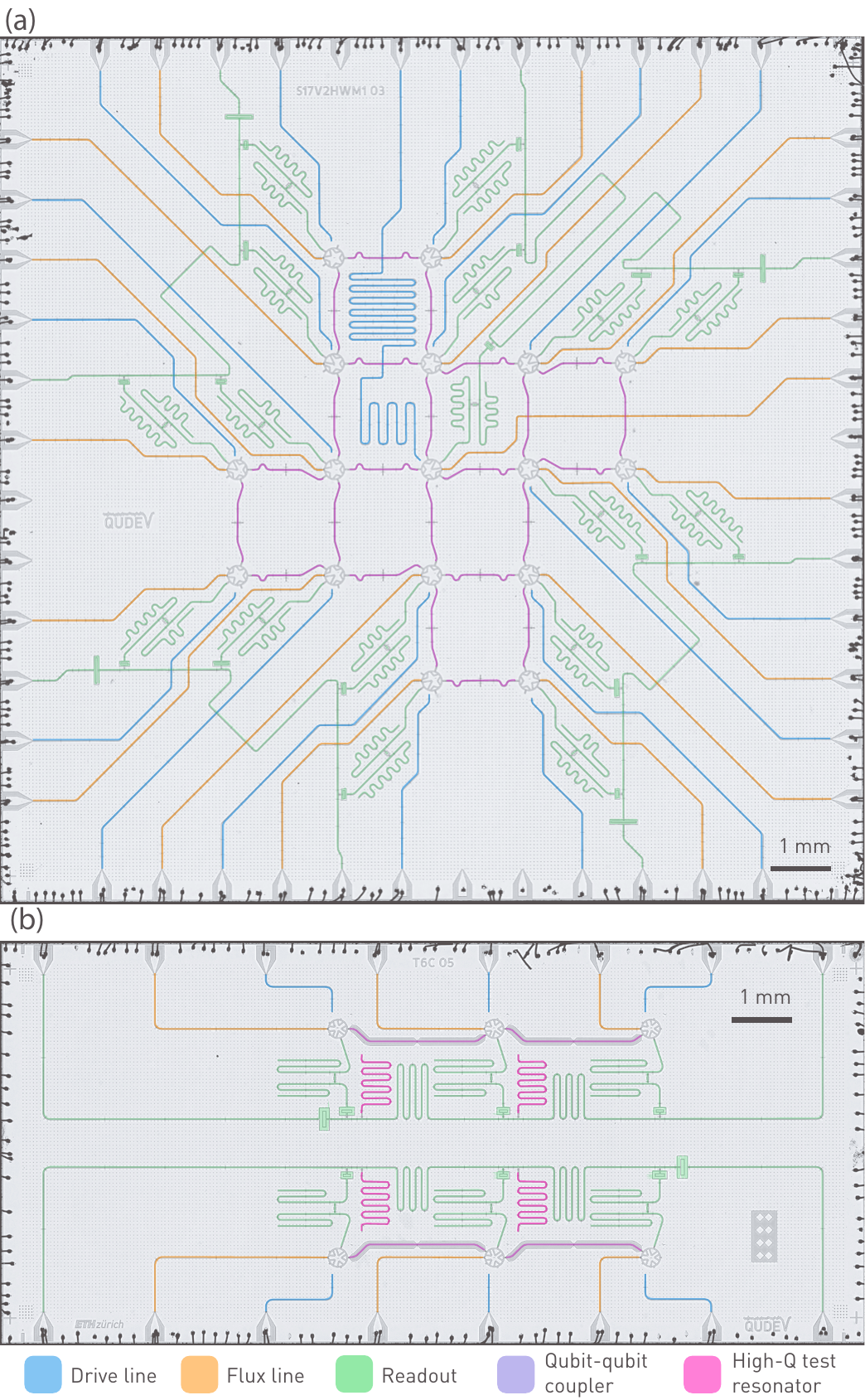}
\caption{Optical micrographs of (a) sample [4-0] with 17 qubits and of (b) [5-0] with 6 qubits from which the data was acquired.}
\label{fig:supplemental-micrographs}
\end{figure}

The fabrication of chip [0-0] is described in~\cite{Krinner2022}. The remaining chips were all fabricated as discussed below. Here, water refers to deionised water and the processes take place at room temperature (RT), unless specified otherwise. 

In the first fabrication step we define the resonators and qubit islands by structuring the \ch{Nb} base metallization layer. We start this process by conditioning an intrinsic high-resistivity (>\,\SI{20}{kOhm}) (100)-silicon wafer in base piranha (30\% \ch{H2O2} 1:1 25\% \ch{NH4}) for \SI{5}{min} at \SI{60}{\celsius} followed by a rinse in water. We then remove oxides on the wafer by immersion in a solution of 7\% hydrofluoric acid (HF) in water for \SI{1}{min}. The reaction is stopped with a rinse in water. We transfer the wafer into a magnetron sputtering tool in which a 125 nm-thick film of Nb is deposited at a pressure of \num{7.5e-3} \SI{}{Torr} with a plasma excitation power of \SI{500}{W}. The device design is patterned onto the wafer in a subtractive process using the lithographic photoresist 
AZ5214E and a \ch{SF6}-only reactive ion-etching with a power of \SI{100}{W} at a pressure of \num{5e-3} \SI{}{Torr}. We strip the resist in a sequence of N-Methyl-2-pyrrolidone (NMP) at \SI{80}{\celsius} for \SI{40}{min} and sonication in solvent baths of NMP, acetone and isopropyl alcohol (IPA) at \SI{50}{\celsius} for \SI{10}{min} each.%

The fabrication of air-bridges follows to connect adjacent ground planes and to create crossovers between control lines. 
We first clean the wafer in a solution of 7\% HF in water for \SI{1}{min} then rinse it in water. 
We pattern the air-bridges in a liftoff process using a bilayer resist stack. The bottom resist layer consists of a copolymer of 8.5\% of methacrylic acid (MAA) in methyl methacrylate (MMA), diluted in ethyl lactate (EL) at a concentration of 12\% [P(MMA-MAA) 12\%]. This layer defines the bases of the air-bridges. We pattern this resist layer using deep ultraviolet (UV) lithography and develop the features in a solution of IPA 1:1 methyl isobutyl ketone (MIBK) for \SI{1}{min}. We proceed to reflow the copolymer resist layer at \SI{200}{\celsius} for \SI{15}{min}. We then spin the top layer of the resist stack, which consists of AZ5214E. We use image reversal photolithography to define the rectangular spans forming the air-bridges. We develop in  AZ 726 MIF for \SI{60}{s}. We install the wafer into an electron-beam evaporator and pump the chamber to a pressure of \num{2e-6} \SI{}{Torr}. We then deposit a trilayer film of 300 nm of Al, 300 nm of Ti and 300 nm of Al at a rate of \SI{0.5}{nm/s}. Liftoff is performed with a pipette in a sequence of warm dimethyl sulfoxide (DMSO) at \SI{80}{\celsius} for \SI{1}{h} and \SI{10}{min} both in acetone and IPA.

We dice the wafer into individual chips for Josephson junction fabrication after protecting the metalization and air-bridges with AZ5214E resist. The protective resist film is stripped in a sequence of \SI{30}{min} of NMP at \SI{80}{\celsius} and sonication in solvent baths at \SI{50}{\celsius} of NMP, acetone and IPA for \SI{5}{min} each. The samples are rinsed in ultrapure IPA and blow dried with \ch{N2}.

The Josephson junctions are defined by spinning a bilayer of \SI{800}{nm} thick of high-sensitivity P(MMA-MAA) 12\% soft baked at \SI{180}{\celsius} for \SI{300}{s}, and \SI{200}{nm} of CSAR 62 (AR-P 6200.9) soft baked at \SI{160}{\celsius} for \SI{240}{s}. The perpendicular Manhattan patterns for the junctions leads are exposed in a \SI{100}{keV} electron-beam lithography (EBL) system 
with doses of \SI{1100}{\micro C / cm ^2} and \SI{170}{\micro C / cm ^2} respectively for the main features and undercut areas. The development is performed in a sequence of \SI{1}{min} of AR 600-546 (n-amyl acetate) at \SI{0}{\celsius}, then \SI{30}{s} of IPA, \SI{10}{min} in MIBK 1:3 IPA and lastly in \SI{30}{s} of IPA.

We then apply the corresponding pre-junction deposition cleaning treatment specified in Tab.~\ref{tab:analyzedchipsummary}. The samples treated with \emph{ion-milling (treatment A)} are installed into the the loadlock of an electro beam evaporator, 
where the ion-milling takes place once for each lead orientation of Josephson junction at a tilt of \ang{45} for \SI{10}{s} with an ion-beam current of \SI{15}{mA} accelerated at \SI{150}{V}. Ion-milling takes place \textit{in-situ} immediately before the deposition of the junctions.
The \emph{ashing treatments B, C1 and C2} were done in an oxygen plasma asher for \SI{10}{s} at \SI{50}{W} and \SI{1}{mbar}.
The samples treated with the additional steps of \emph{BHF} (buffered hydrofluoric acid) were dipped into a solution of \ch{NH4F} 7:1 \ch{HF} for the corresponding duration of either \SI{1}{min} \emph{(treatment C1)} or \SI{10}{min} \emph{(treatment C2)}. After the BHF treatment, we rinse the samples in water for \SI{1}{min} to stop the reaction.  

For the deposition of the junctions, the ion-milled samples (treatment A) remain in the electron beam deposition tool without breaking the vacuum. The remaining samples (treatments B, C1, C2) are installed within \SI{5}{min} of the end of cleaning treatment into the deposition tool. 
The chamber is pumped to a pressure below \num{1e-7} \SI{}{Torr}. 
We tilt the sample holder at an angle of \ang{45} and rotate it to a planetary angle such that Al can only be deposited on the first lead of the Josephson junctions. 
A \SI{30}{nm} film of Al is evaporated at a rate of \SI{0.5}{nm/s}. The samples move to an attached oxidation chamber where the exposed Al film is oxidized in a mixture of 15\% \ch{O2} in Ar under the conditions specified in Tab.~\ref{tab:analyzedchipsummary}. This oxidation step defines the tunnel barrier of the Josephson junctions. The samples are transferred back into the evaporation chamber. There, we tilt the sample holder to an angle of \ang{45}, but this time, we rotate it to a planetary rotation different by \ang{90} to that of the first  junction lead. This way, only the features corresponding to the second lead of the Josephson junctions are exposed to the deposition. We then evaporate a \SI{80}{nm}-thick Al film at a rate of \SI{0.5}{nm/s}. After deposition, the Al film is oxidized in the oxidation chamber for \SI{5}{min} at \SI{15}{Torr}. We remove the samples from the electron beam deposition tool and proceed with the liftoff of the deposited film. This is performed with a pipette after \SI{2}{h} in acetone at \SI{50}{\celsius}. The sample is then cleaned in a sequence of \SI{30}{min} of NMP at \SI{80}{\celsius} and sonication in NMP, acetone and IPA at \SI{50}{\celsius} for \SI{5}{min} each. The samples are then rinsed in ultrapure IPA and blow dried with \ch{N2}.

To ensure a superconducting connection between the Josephson junction leads and the base-layer metallization, we fabricate bandages~\cite{Dunsworth2017}. 
For that, we spin on the samples a bilayer resist stack consisting of a \SI{600}{nm}-thick bottom layer of copolymer P(MMA/MAA) 12\% in EL soft baked at \SI{160}{\celsius} for \SI{240}{s} and a top layer of \SI{200}{nm}-thick top layer of PMMA 950 K 4\% in EL baked at \SI{160}{\celsius} for \SI{240}{s}. 
We also spin a discharge layer of Espacer 300Z baked at \SI{80}{\celsius} for \SI{90}{s}. 
We expose the samples in a \SI{30}{keV} EBL system where the undercut dose is of \SI{60}{\micro C / cm^2} and the main bandage dose is of \SI{400}{\micro C / cm^2}. After exposure, we remove the discharge layer from the samples by dipping them in water for \SI{30}{s}. We then develop the samples in \ch{MIBK} 3:1 \ch{IPA} for \SI{50}{s}. We stop the reaction by placing the samples into IPA for \SI{30}{s}. We then install the samples into an electron beam deposition tool where we perform ion-milling on the bandage openings with a current of \SI{10}{mA} and acceleration voltage of \SI{400}{V} at a tilt of \ang{30} for \SI{1}{min} \SI{30}{s} with constant planetary motion of \SI{4}{RPM}. Then we deposit \SI{300}{nm} of Al at a rate of \SI{0.5}{nm/s}, at \ang{10} tilt with constant planetary rotation of \SI{4}{RPM}. After the deposition, the exposed Al is controllably oxidized for \SI{10}{min} at \SI{10}{Torr}. The deposited film is lifted off after a \SI{2}{h} solvent bath in acetone at \SI{50}{\celsius} with a pipette. The samples are then cleaned in a sequence of \SI{30}{min} of NMP at \SI{80}{\celsius} and sonication in NMP, acetone and IPA at \SI{50}{\celsius} for \SI{5}{min} each. The samples are then rinsed in ultrapure IPA and blow dried with \ch{N2}.

We spin a thick layer of AZ5214E and dice the sample, this time to the dimensions of the printed circuit board (PCB) openings. The protective resist layer is removed in a sequence of solvent baths. First in NMP at \SI{80}{\celsius}, and then with sonication in solvent baths of NMP, acetone and IPA, all at \SI{50}{\celsius} for \SI{5}{min} each. We then rinse the samples in ultrapure IPA and blow dry them. After thorough inspection, the samples are glued onto the sample mount and we electrically connect the control lines and ground plane of the sample to the PCB with aluminum bonds. Lastly, we install the samples into the dilution refrigerator. Exemplary micrographs of samples with 17 and 6 qubits are displayed in Fig.~\ref{fig:supplemental-micrographs}.

\section{Table of devices}
\label{sec:supplemental-table}

We describe in Tab.~\ref{tab:analyzedchipsummary} the investigated samples. The first and second digits of the chip label indicate the wafer and sample numbers respectively.


\begin{table*}[]
\caption{List of devices investigated for this work.}
\label{tab:analyzedchipsummary}
\begin{tabular}{ccccccccc}
\hline
\multicolumn{1}{c|}{Sample}                                                      & \multicolumn{3}{c|}{Fabrication}                                                                                                                                                                                 & \multicolumn{3}{c|}{Qubit properties}                                                                                                                                                                                                                & \multicolumn{2}{c}{Swap-spectroscopy}                                                                                                    \\ \cline{2-9} 
\multicolumn{1}{c|}{\begin{tabular}[c]{@{}c@{}}label\\ (wf.-ch.) \end{tabular}} & \multicolumn{1}{c|}{Cleaning} & \multicolumn{1}{c|}{\begin{tabular}[c]{@{}c@{}}$j_\mathrm{c}$ \\ (\SI{}{\micro \A \per \micro \m \squared})\end{tabular}} & \multicolumn{1}{c|}{\begin{tabular}[c]{@{}c@{}}Oxidation \\ parameters\end{tabular}} & \multicolumn{1}{c|}{\begin{tabular}[c]{@{}c@{}}no. \\ qubits\end{tabular}} & \multicolumn{1}{c|}{\begin{tabular}[c]{@{}c@{}}$\omega_{\mathrm{qb}}$\\  (GHz)\end{tabular}} & \multicolumn{1}{c|}{\begin{tabular}[c]{@{}c@{}}$T_{1}$\\  (\SI{}{\micro \s})\end{tabular}} & \multicolumn{1}{c|}{\begin{tabular}[c]{@{}c@{}}no. \\ defects\end{tabular}} & \begin{tabular}[c]{@{}c@{}}Bandwidth\\  (GHz)\end{tabular} \\ \hline
0-0~\cite{Krinner2022}                                                                            & A: ion-milling                   & 0.6                                                                                       & 12 min, 10 Torr                                                                      & 17                                                                         & 5.32 $\pm$ 0.75                                                                     & 32.5 $\pm$ 16.3                                                                   & 24.3 $\pm$ 3.4                                                              & 27.95                                                      \\
1-0                                                                            & A: ion-milling                   & 1.6                                                                                       & 3 min, 3 Torr                                                                        & 2                                                                          & 6.74 $\pm$ 0.05                                                                     & 10.9 $\pm$ 0.7                                                                    & 1                                                                           & 2.27                                                       \\
1-1                                                                            & A: ion-milling                   & 1.6                                                                                       & 3 min, 3 Torr                                                                        & 5                                                                          & 6.64 $\pm$ 0.27                                                                     & 12.1 $\pm$ 5.1                                                                    & 1                                                                           & 4.78                                                       \\
1-2                                                                            & A: ion-milling                   & 1.6                                                                                       & 3 min, 3 Torr                                                                        & 2                                                                          & 6.64 $\pm$ 0.20                                                                     & 14.2 $\pm$ 6.4                                                                    & 3                                                                           & 2.39                                                       \\
1-3                                                                            & A: ion-milling                   & 1.6                                                                                       & 3 min, 3 Torr                                                                        & 3                                                                          & 6.45 $\pm$ 0.23                                                                     & 10.1 $\pm$ 4.3                                                                    & 2                                                                           & 3.65                                                       \\
1-4                                                                            & A: ion-milling                   & 1.6                                                                                       & 3 min, 3 Torr                                                                        & 3                                                                          & 6.67 $\pm$ 0.10                                                                     & 14.0 $\pm$ 6.1                                                                    & 0                                                                           & 1.24                                                       \\
1-5                                                                            & A: ion-milling                   & 1.6                                                                                       & 3 min, 3 Torr                                                                        & 2                                                                          & 7.09 $\pm$ 0.46                                                                     & 2.7 $\pm$ 1.7                                                                     & 3                                                                           & 3.86                                                       \\
2-0                                                                            & C2: ashing + 10 min BHF           & 1.6                                                                                       & 3 min, 3 Torr                                                                        & 0                                                                          & -                                                                                   & -                                                                                 & 0                                                                           & 0                                                          \\
2-1                                                                            & C2: ashing + 10 min BHF           & 1.6                                                                                       & 3 min, 3 Torr                                                                        & 5                                                                          & 6.28 $\pm$ 0.18                                                                     & 19.8 $\pm$ 6.7                                                                    & 6                                                                           & 19.93                                                      \\
2-2                                                                            & C2: ashing + 10 min BHF           & 1.6                                                                                       & 3 min, 3 Torr                                                                        & 0                                                                          & -                                                                                   & -                                                                                 & 0                                                                           & 0                                                          \\
2-3                                                                            & C2: ashing + 10 min BHF           & 1.6                                                                                       & 3 min, 3 Torr                                                                        & 5                                                                          & 6.35 $\pm$ 0.30                                                                     & 19.7 $\pm$ 5.6                                                                    & 0                                                                           & 0                                                          \\
2-4                                                                            & C1: ashing + 1 min BHF            & 1.6                                                                                       & 3 min, 3 Torr                                                                        & 4                                                                          & 4.85 $\pm$ 0.26                                                                     & 44.3 $\pm$ 14.2                                                                   & 5                                                                           & 8.41                                                       \\
2-5                                                                            & C1: ashing + 1 min BHF            & 1.6                                                                                       & 3 min, 3 Torr                                                                        & 6                                                                          & 4.80 $\pm$ 0.14                                                                     & 41.2 $\pm$ 10.3                                                                   & 3                                                                           & 11.95                                                      \\
3-0                                                                            & B: ashing                        & 1.6                                                                                       & 3 min, 3 Torr                                                                        & 6                                                                          & 5.92 $\pm$ 0.20                                                                     & 39.8 $\pm$ 7.7                                                                    & 10                                                                          & 16.66                                                      \\
3-1                                                                            & B: ashing                        & 1.6                                                                                       & 3 min, 3 Torr                                                                        & 6                                                                          & 6.40 $\pm$ 0.28                                                                     & 28.6 $\pm$ 8.5                                                                    & 4                                                                           & 14.46                                                      \\
4-0                                                                            & C1: ashing + 1 min BHF            & 0.6                                                                                       & 12 min, 10 Torr                                                                      & 17                                                                         & 3.89 $\pm$ 0.67                                                                     & 44.0 $\pm$ 29.5                                                                   & 17                                                                          & 14.87                                                      \\
5-0                                                                            & B: ashing                        & 0.3                                                                                       & 60 min, 50 Torr                                                                      & 4                                                                          & 4.81 $\pm$ 0.25                                                                     & 10.0 $\pm$ 9.2                                                                    & 11                                                                          & 6.18                                                       \\
5-1                                                                            & B: ashing                        & 0.3                                                                                       & 60 min, 50 Torr                                                                      & 5                                                                          & 5.33 $\pm$ 0.19                                                                     & 14.80 $\pm$ 11.83                                                                 & 19                                                                          & 10.58                                                      \\ \hline
\textbf{$\Sigma$}                                                                       &                               &                                                                                           &                                                                                      & 92                                                                         &                                                                                     &                                                                                   & 109.3  $\pm$ 3.4                                                            & 149.18                                                     \\ \hline
\end{tabular}
\end{table*}


\section{Counting threshold sweep}
\label{sec:supplemental-threshold}

\begin{figure}[H]
\centering
\includegraphics[width=1\linewidth]{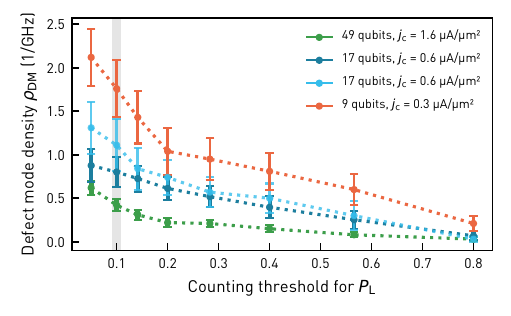}
\caption{Defect mode densities $\rho_{\mathrm{DM}}$ versus counting thresholds for $P_\mathrm{L}$. }
\label{fig:supplemental-threshold}
\end{figure}

\begin{figure}[H]
\centering
\includegraphics[width=1\linewidth]{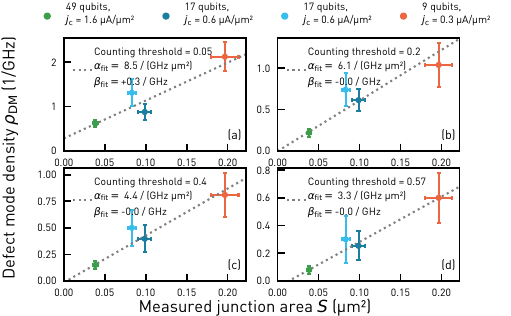}
\caption{$\rho_{\mathrm{DM}}$ versus junction sizes for four different counting thresholds: (a) 0.05, (b) 0.2, (c) 0.4 and (d) 0.57. The dotted lines correspond to a fit of the line $\rho_{\mathrm{DM}} = \alpha_{\mathrm{fit}} \, S + \beta_{\mathrm{fit}}$.}
\label{fig:supplemental-threshold-fourplots}
\end{figure}

The spectral density of strongly coupled defect modes, $\rho_{\mathrm{DM}}$, is sensitive to measurement properties such as interaction time $\tau$ but also to parameters used in the data analysis, such as the counting threshold. 
The counting threshold corresponds to the minimal population loss $P_\mathrm{L}$ induced by a defect so that it is counted as a strongly coupled defect.
As we make the counting threshold higher, the number of strongly coupled defect modes decreases because of the stricter counting requirements  (see Fig.~\ref{fig:supplemental-threshold}). We note, however, that the increasing trend of $\rho_{\mathrm{DM}}$ with increasing junction area from Fig.~\ref{fig:fig3}(b) is preserved for any chosen counting threshold. Additionally, the linearity of the trend seems to be preserved as well, as shown in Fig.~\ref{fig:supplemental-threshold-fourplots}.

\section{Procedure for the defect analysis}
\label{sec:supplemental-algorithm}

Our peak-finding algorithm takes as an input a dataset of $\ket{e}$-state population loss ($P_\mathrm{L}$) versus frequency. We start by identifying possible candidates for defect frequencies. For that, we smoothen the data by applying a second-order Savitzky-Golay (SG) filter, which preserves the linewidth of the defect signatures. This represents a balanced choice, in comparison to a zeroth-order SG filter (boxcar), which excessively smoothens the signal, and a higher-order (\textit{i.e.} fourth-order) SG filter, which can lead to the undesired retention of spurious peaks originating from coherent defect-qubit energy exchange. For our analysis the SG filtering window was set to \SI{143}{MHz}.

We proceed to identify all frequencies in the SG-filtered data that correspond to local maxima with prominence of at least $0.03$. The maxima are counted as defects if the raw $P_\mathrm{L}$ at the frequency is above the counting threshold and there are no other defects within 100 MHz of it. The latter condition discards data stemming from coherent oscillations in $P_\mathrm{L}$ to a single strongly coupled defect via exchange of population.         

To estimate the uncertainty of the spectral density of strongly coupled defect modes $\rho_{\mathrm{DM}}$ of a collection of defect mode data from a set of qubits or a set of samples, we perform bootstrapping~\cite{Arch2024} utilizing a sufficiently high number of repartitions and resampling repetitions. The reported spectral defect mode densities $\rho_{\mathrm{DM}}$ correspond to the bootstrap estimator of the distribution mean while the reported uncertainties correspond to the 68\% confidence intervals for $\rho_{\mathrm{DM}}$. 

\section{Experimental setup}
\begin{figure}[H]
\centering
\includegraphics[width=1\linewidth]{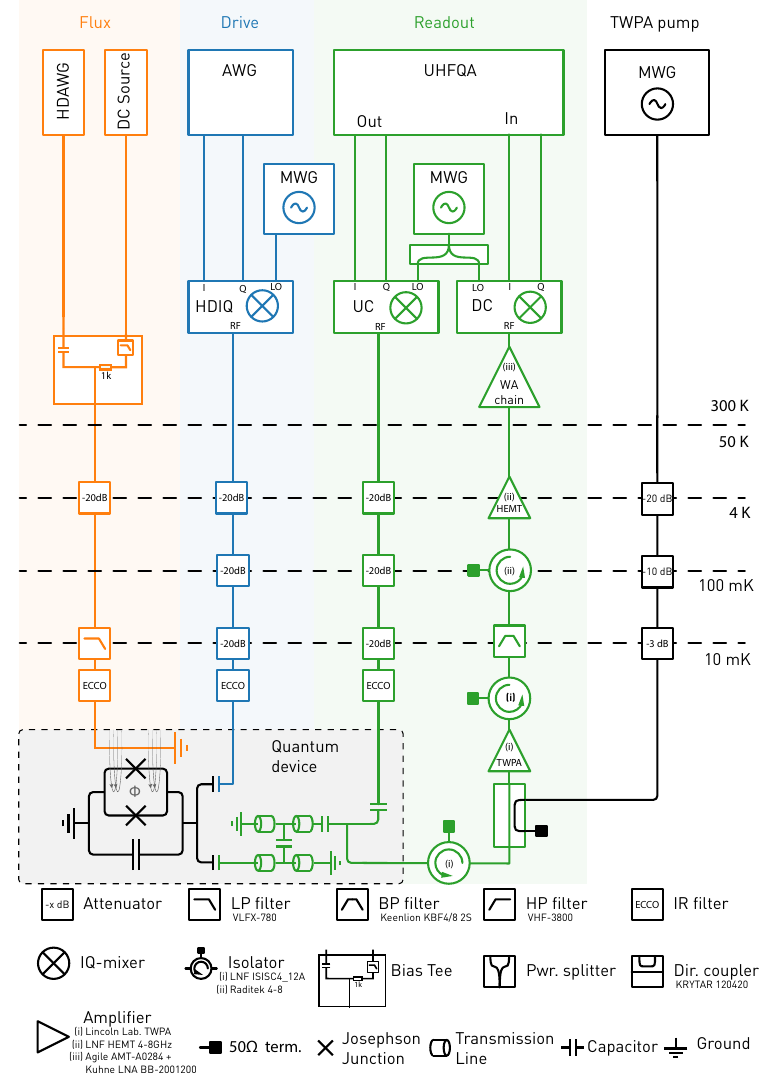}
\caption{Wiring diagram.}
\label{fig:diagramcabeling}
\end{figure}

The experimental setup utilized to perform the measurements is represented in Fig.~\ref{fig:diagramcabeling}.
\\

\end{appendix}

\bibliography{references}

\end{document}